\documentclass[amsmath,amssymb,aps,twocolumn,prb,superscriptaddress]{revtex4-1}

\usepackage[usenames]{xcolor}
\usepackage[colorlinks=true,citecolor=blue,linkcolor=magenta]{hyperref}
\usepackage{filecontents}

\usepackage{titlesec}
\definecolor{natureblue}{HTML}{343796}
\titleformat{\section}[display]{\vspace{-1em}}{}{0pt}{\normalfont\sffamily\color{natureblue}}[\vspace{-3pt}\hrule]
\usepackage{setspace}
\makeatletter
\def\frontmatter@abstractfont{\sffamily\color{natureblue}\setstretch{1.5}}%
\def\frontmatter@title@format{\noindent\huge\sffamily}{}%
\def\frontmatter@authorformat{\vspace{1em}\noindent\color{natureblue}\Large\sffamily}%
\def\frontmatter@affiliationfont{\vspace{1em}\color{black}\noindent\normalsize\sffamily}%
\def\frontmatter@above@affiliation@script{\vspace{1em}\noindent}%
\def\frontmatter@makefnmark{}
\renewcommand*\frontmatter@date[2][\Dated@name]{\def\@date{}}%
\makeatother

\usepackage{amsmath,amsfonts,amssymb,amsthm,graphics,graphicx,times,bbm}

\newcommand{\tr}{\mathrm{tr}}

\usepackage{cancel}

\begin{document}

\title{Advances in Quantum Teleportation}

\begin{abstract}
Quantum teleportation is one of the most important protocols in
quantum information. By exploiting the physical resource of
entanglement, quantum teleportation serves as a key primitive in a
variety of quantum information tasks and represents an important
building block for quantum technologies, with a pivotal role in
the continuing progress of quantum communication, quantum
computing and quantum networks. Here we review the basic
theoretical ideas behind quantum teleportation and its variant
protocols. We focus on the main experiments, together with the
technical advantages and disadvantages associated with the use of
the various technologies, from photonic qubits and optical modes
to atomic ensembles, trapped atoms, and solid-state systems.
Analysing the current state-of-the-art, we finish by discussing
open issues, challenges and potential future implementations.
\end{abstract}

\author{S. Pirandola}
\affiliation{\small 1 Computer Science and York Centre for Quantum
Technologies, University of York, York YO10 5GH, United Kingdom}
\author{J. Eisert}
\affiliation{\small 2 Dahlem Center for Complex Quantum Systems,
Freie Universit\"{a}t Berlin, 14195 Berlin, Germany}
\author{C. Weedbrook}
\affiliation{\small 3 Department of Physics, University of
Toronto, Toronto M5S 3G4, Canada}
\author{A. Furusawa}
\affiliation{\small 4 Department of Applied Physics, The
University of Tokyo, 7-3-1 Hongo, Bunkyo-ku, Tokyo 113-8656,
Japan}
\author{S. L. Braunstein}
\affiliation{\small 1 Computer Science and York Centre for Quantum
Technologies, University of York, York YO10 5GH, United Kingdom}


\maketitle

\section{From Science Fiction to Reality}

It has been over two decades since the discovery of quantum
teleportation, in what is arguably one of the most interesting and
exciting implications of the `weirdness' of quantum mechanics.
Previous to this landmark discovery, this fascinating idea
belonged to the realm of science fiction. First coined in a 1931
book by Charles H. Fort~\cite{LO31}, the term teleportation has
since been used to refer to the process by which bodies and
objects are transferred from one location to another, without
actually making the journey along the way. Since then it has
become a fixture of pop culture, perhaps best exemplified by Star
Trek's celebrated catchphrase \textquotedblleft Beam me up,
Scotty.\textquotedblright

In 1993, a seminal paper~\cite{Bennett93} described a quantum
information protocol, dubbed quantum teleportation, that shares
several of the above features. In this protocol, an unknown
quantum state of a physical system is measured and subsequently
reconstructed or `reassembled' at a remote location (the physical
constituents of the original system remain at the sending
location). This process requires classical communication and
excludes superluminal communication. Most importantly, it requires
the resource of quantum entanglement \cite{HoroRMP,OldReview}.
Indeed, quantum teleportation can be seen as the protocol in
quantum information that most clearly demonstrates the character
of quantum entanglement as a resource: Without its presence, such
a quantum state transfer would not be possible within the laws of
quantum mechanics.

Quantum teleportation plays an active role in the progress of
quantum information
science~\cite{NielsenChuang,Wilde,WeeRMP,BraRMP}. On the one hand
it is a conceptual protocol crucial in the development of formal
quantum information theory, on the other it represents a
fundamental ingredient to the development of many quantum
technologies. Schemes such as quantum repeaters~\cite{Briegel98}
-- pivotal for quantum communication over large distances --
quantum gate teleportation~\cite{Gottesman99}, measurement-based
computing~\cite{Raussendorf01}, and port-based
teleportation~\cite{Programmable} all derive from the basic
scheme. The vision of a quantum network~\cite{Kimble2008} draws
inspiration from it. Teleportation has also been used as a simple
tool for exploring `extreme' physics, such as closed time-like
curves~\cite{Lloyd11} and black-hole evaporation~\cite{Lloyd14}.

Today, quantum teleportation has been achieved in laboratories the
world over using many different substrates and technologies,
including photonic qubits (e.g., light
polarisation~\cite{Bouwmeester97,Boschi98,Ursin04,Jin10,Kim01,Yin12,Ma12},
time-bin~\cite{Marcikic03,deRiedmatten04,Landry07}, dual rails on
chip~\cite{Metcalf14}, spin-orbital qubits~\cite{Wang15}), nuclear
magnetic resonance (NMR)~\cite{Nielsen98}, optical
modes~\cite{Furusawa98,Bowen03,Zhang03,Takei05,Takei05b,Yonezawa07,Lee11,Yukawa08,Takeda13}
, atomic ensembles~\cite{Sherson06,Krauter13,Chen08,Bao12},
trapped
atoms~\cite{Riebe04,Riebe07,Barrett04,Olmschenk09,Nolleke13}, and
solid state systems~\cite{Bussieres14,Gao13,Steffen13,Pfaff14}.
Outstanding performances have been achieved in terms of
teleportation distance~\cite{Yin12,Ma12}, with satellite-based
implementations forthcoming. Attempts at scaling to more complex
quantum systems have also been started~\cite{Wang15}.

\begin{figure*}[t!]
\vspace{-0.4cm}
\par
\begin{center}
\includegraphics[width=0.99\textwidth] {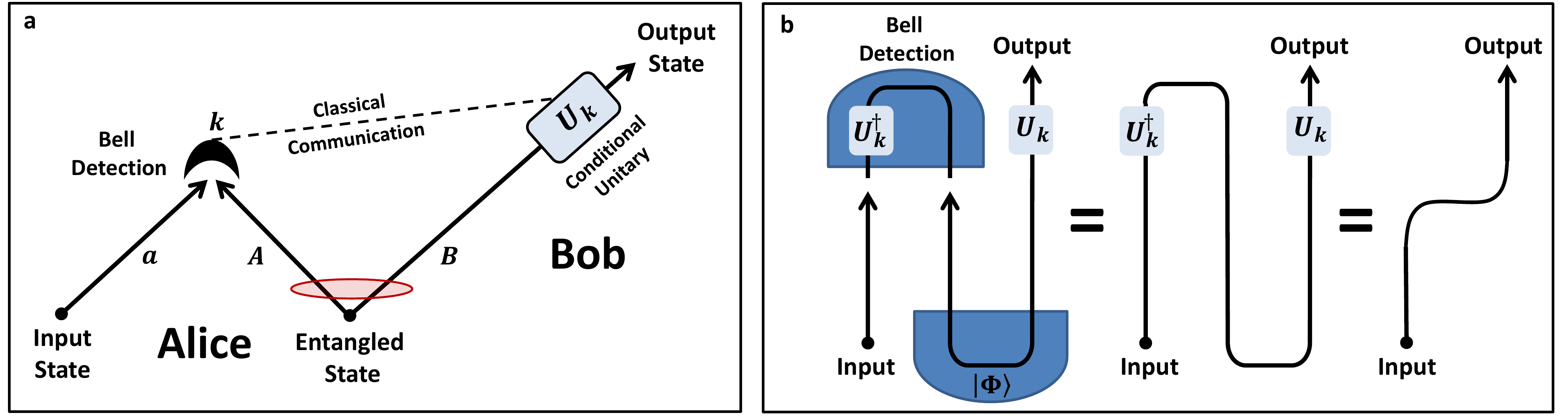}
\end{center}
\par
\vspace{-0.4cm} \caption{Panel~\textbf{a} shows the basic protocol
of quantum teleportation. Alice's unknown input state is
teleported to Bob using a shared entangled state and a classical
communication channel. Alice performs a Bell state measurement on
her systems, $a$ and $A$, and classically communicates the outcome
$k$ to Bob. Using the measurement result, Bob applies the
conditional unitary $U_k$ to his system $B$, retrieving an output
state which is an exact replica of Alice's input state in the
ideal case ($F=1$). The outcomes of the Bell detection are:
$k=0,\dots, d^2-1$ for DV systems, and $k\in\mathbb{C}$ for
continuous-variable (CV) systems. In qubit-teleportation ($d=2$),
the input state is typically pure, the entangled state is a Bell
pair, and $U_k$ is a qubit Pauli operator ${\cal P}_k$. In CV
teleportation ($d=\infty$), the systems are bosonic modes (e.g.,
optical), the input is typically a coherent state, the entangled
state is an EPR state, and $U_k$ is a phase-space displacement.
Variations on this basic protocol may occur depending on the
actual technologies adopted. Panel~\textbf{b} is a representation
of the ideal teleportation process in terms of Penrose-inspired
space-time diagrams~\cite{BraunsteinTwist00,Abramsky04}.
Let us consider a maximally entangled state $\vert \Phi \rangle$
as a shared quantum resource (i.e., a Bell pair for qubits, or an
ideal EPR state for CVs). Then, the Bell detection corresponds to
applying $\langle \Phi \vert (U_k^{\dagger} \otimes I)$ where the
outcome $k$ is randomly selected by the measurement. As a result,
an arbitrary input state $\rho$ is transformed into $U_k^{\dagger}
\rho U_k$ where the unitary $U_k^\dagger$ is correspondingly
`undone' by Bob.}\label{pic1} 
\end{figure*}

\section{The Basics of Quantum Teleportation}

\subsection{Quantum Teleportation of Qubits and Discrete Variables}

Quantum teleportation was originally described for two-level
quantum systems, so-called qubits~\cite{Bennett93}. The protocol
considers two remote parties, referred to as Alice and Bob, who
share two qubits, $A$ and $B$, prepared in a pure entangled state.
In the ideal case, this is taken to be maximally entangled, such
as $|\Phi\rangle=(|0,0\rangle +|1,1\rangle)/\sqrt{2}$, also known
as a Bell pair~\cite{NielsenChuang}. At the input, Alice is given
another qubit $a$ whose state $\rho$ is unknown. She then performs
a joint quantum measurement, called Bell
detection~\cite{Weinfurter94,Braunstein95}, which projects her
qubits $a$ and $A$ into one of the four Bell states
$(\mathcal{P}_{k}\otimes I)|\Phi \rangle$ with $k=0,\dots,3$ where
$\mathcal{P}_{k}$ is a qubit Pauli
operator~\cite{NielsenChuang,Wilde}. As a result, the state of
Alice's input qubit has been collapsed by measurement while Bob's
qubit $B$ is simultaneously projected onto
$\mathcal{P}^\dagger_{k} \rho \mathcal{P}_{k}$. In the last
(feed-forward) step of the protocol, Alice communicates the
classical outcome $k$ of her measurement to Bob, who then applies
$\mathcal{P}_{k}$ to recover the original input state $\rho$.

Note that Alice's input state is assumed to be unknown, otherwise
the protocol reduces to remote state preparation~\cite{Bennett01}.
Furthermore, this state may itself be part of a larger composite
quantum system shared with a third party (in which case,
successful teleportation requires reproducing all correlations
with that third party). In typical experiments, the input state is
been taken to be pure and belonging to a limited alphabet of
states, e.g., the six poles of the Bloch
sphere~\cite{NielsenChuang}. 
 In the presence of decoherence, the
quality of the reconstructed state may be quantified by its
teleportation fidelity, $F\in\lbrack0,1]$. 
This is the fidelity~\cite{NielsenChuang} between
Alice's input state and Bob's output state, averaged over all the
outcomes of the Bell
detection and input state alphabet. 
For small values of the fidelity, strategies exist that allow for
an imperfect teleportation while making no use of any entangled
resource. For example, Alice may directly measure her input state,
sending the results to Bob for him to prepare an output
state\cite{Scarani05}. Such a measure-prepare strategy is called
`classical teleportation' and has the maximum
fidelity~\cite{Massar95} $F_{\text{class}}=2/3$ for an arbitrary
input state or, equivalently, an alphabet of mutually unbiased
states, such as the six poles of the Bloch sphere. Thus, one clear
benchmark to ensure that quantum resources are utilized is to
require fidelity $F>F_{\text{class}}$.

Quantum teleportation is not restricted to qubits, but may involve
higher-dimensional quantum systems. For every finite dimension
$d$, one can formulate ideal teleportation schemes: This exploits
a basis of maximally entangled state vectors, which can be derived
from a given maximally entangled state, and a basis $\{U_k\}$ of
unitary operators satisfying~\cite{AllTeleportation}
$\tr(U_j^\dagger U_k ) = d\, \delta_{j,k}$. The scheme follows
then the above lines, with the Pauli operators ${\cal P}_k$
replaced by $U_k$ and the classical channel involves
distinguishing $d^2$ signals. Such a protocol can be constructed
for any finite-dimensional Hilbert space, for so-called
discrete-variable (DV) systems. See Fig.~\ref{pic1}.

\subsection{Quantum Teleportation of Continuous Variables}

Quantum teleportation can also be extended to quantum systems with
an infinite-dimensional Hilbert space, known as
continuous-variable (CV) systems. These are typically realised by
optical bosonic modes, whose electric field can be described by
position- and momentum-like quadrature
operators~\cite{BraRMP,WeeRMP}. Following the first theoretical
proposals~\cite{Vaidman94,Braunstein98}
, CV teleportation was demonstrated with optical
modes~\cite{Furusawa98}. Other CV systems can be considered,
including opto-mechanical systems~\cite{Pirandola03,Mancini03} and
collective spins of atomic
ensembles~\cite{Sherson06,Krauter13,Kuzmich00,Duan00,Hammerer10}.

In the standard CV protocol with optical modes, the entangled
resource corresponds to a two-mode squeezed vacuum
state~\cite{BraRMP}, also known as an Einstein-Podolsky-Rosen
(EPR) state. This is a zero-mean Gaussian state~\cite{WeeRMP} with
logarithmic negativity~\cite{PhD,Vidal02,Plenio05}
$-\log(\varepsilon)$, where the parameter $\varepsilon\leq1$ can
be computed from its covariance matrix~\cite{WeeRMP}. Suppose
Alice is given an input mode $a$ in an unknown state; typically, a
coherent state $\left\vert \alpha\right\rangle $ with unknown
amplitude $\alpha$. She then applies a CV Bell measurement on her
modes, $a$ and $A$, which consists of mixing them on a balanced
beam splitter and homodyning the output ports in the conjugate
quadratures $\hat{q}_{-}=(\hat{q}_{a}-\hat{q}_{A})/\sqrt{2}$ and $\hat{p}%
_{+}=(\hat{p}_{a}+\hat{p}_{A})/\sqrt{2}$. In the last step, Alice
communicates the classical outcome $k=q_{-}+ip_{+}$ to Bob, who
performs a conditional displacement on his mode $B$ (see
Fig.~\ref{pic1})

The teleportation fidelity for the alphabet of coherent states is
given by $F=(1+\varepsilon)^{-1}$. Perfect teleportation occurs
only for unbounded entanglement $\varepsilon\rightarrow0$, where
the entangled state approximates the ideal EPR state, realising
the perfect correlations $\hat{q}_{A}=\hat{q}_{B}$ and
$\hat{p}_{A}=-\hat{p}_{B}$. Without entanglement ($\varepsilon=1$)
we have $F=1/2$, which is the classical (i.e., no entanglement)
threshold for teleporting coherent
states~\cite{Braunstein01,Hammerer05}. A more stringent threshold
asks that the teleported state is the best copy of the input
allowed by the no-cloning bound~\cite{Grosshans01}, requiring
$F>2/3$. This basic CV protocol can be extended to more general
Gaussian state
settings~\cite{Pirandola06,Fiurasek02,Pirandola05,Pirandola05b,Adesso08,Chiribella14}.

\section{Variants of Teleportation}

Teleportation is an important primitive which has been extended in
various ways. Some of these extensions are actual protocols of
quantum technology, while others are of conceptual value in
theoretical models. In this section, we mention some of these
variants and sketch some recent developments.

\subsection{Entanglement Swapping and Quantum Repeaters}

As previously mentioned, the input to be teleported can itself be
part of an entangled state. Teleportation would then also transfer
the entanglement. When the Bell detection is performed by a third
party, say Charlie acting as the middle relay, this variant is
called entanglement swapping. Suppose that Alice and Bob do not
share entanglement, but they locally prepare two entangled states;
$\rho_{aA}$ at Alice's station and $\rho_{bB}$ at Bob's. They
retain systems $a$ and $b$, while sending systems $A$ and $B$ to
Charlie for Bell detection. After the measurement outcome $k$ is
classically broadcast, Alice and Bob share the conditional output
state $\rho_{ab|k}$ which is entangled. This can be implemented
with DV systems~\cite{Zukowski93} (e.g., polarisation
qubits~\cite{Pan98,Pan01,Jennewein02}), CV
systems~\cite{vanLoock99,Polkinghorne99,Tan99,Johnson02,Pirandola06b,Abdi12}
(e.g., optical modes~\cite{Takei05,Jia04}), or adopting a hybrid
approach involving both DVs and CVs~\cite{Takeda15}.

This protocol forms the basis for quantum
repeaters~\cite{Briegel98}, in which the combination of
entanglement swapping and distillation~\cite{Bennett96,Eisert04}
allows for the distribution of entanglement over large distances.
Once that maximal entanglement is distilled along a long chain of
repeaters, teleportation between the end-users provides the
perfect transfer of quantum information. The remote correlations
created by swapping can also be exploited in quantum cryptography:
Even when Charlie is present in the role of an eavesdropper
(untrusted relay), Alice and Bob can transform their resulting
correlations into a secret key~\cite{Braunstein12}.

\subsection{Quantum Teleportation Networks}

Another important extension is to a quantum teleportation network,
where $n>2$ parties initially share a multipartite entangled state
and teleportation can be performed between any two parties. For
simplicity, we describe the simplest case of a 3-party network,
where, for example, Alice can teleport to either Bob or Charlie.
One simple strategy is assisted teleportation, where Charlie
performs a local measurement on his system and broadcasts the
result with the aim of improving the fidelity of quantum
teleportation from Alice to Bob. This requires Charlie's operation
to be tailored so as to increase the bipartite entanglement of the
remaining parties.

With qubits, 3-party assisted networks can be constructed using
the Greenberger-Horne-Zeilinger state~\cite{Karlsson98}. In
particular, this state makes teleportation equivalent to quantum
secret sharing~\cite{Hillery99}, where Alice's quantum information
can be recovered by Bob if and only if Charlie assists. With CV
systems, 3-party assisted networks can be constructed using
Gaussian states~\cite{Pirandola06}, e.g., generated using squeezed
vacua at the input of two beamsplitters~\cite{vanLoock00} as used
in the first experimental
demonstrations~\cite{Yonezawa04,Lance04}. In general, by using one
or more squeezed vacua in an interferometer, one can create
multi-party assisted networks involving an arbitrary number of
bosonic modes~\cite{vanLoock00,BraRMP}.

A second strategy is unassisted teleportation, where Charlie
rather than helping Alice, also receives a copy of the input. In
this case, the protocol corresponds to quantum telecloning, with
Alice teleporting to Bob and Charlie simultaneously, though with a
teleportation fidelity limited by the no-cloning bound (which is
$5/6$ for qubits~\cite{Buzek96,Bru98}, and $2/3$ for coherent
states of CV systems~\cite{Cerf00}). Quantum telecloning, from one
sender to two recipients, has been experimentally implemented with
polarised photonic qubits~\cite{Zhao05} and coherent states of
optical modes~\cite{Koike06}. Theoretically, the protocol can be
formulated for an arbitrary number of recipients, both in the case
of qubits~\cite{Murao99} and CV systems~\cite{vanLoock01}.

\subsection{Quantum Gate Teleportation and Quantum Computing}

Quantum teleportation can be expressed in terms of primitive
quantum computational operations~\cite{Brassard98} and its
protocol can be extended to quantum gate
teleportation~\cite{Gottesman99,Measurements}. This idea is rooted
in the observation that unitary state manipulation can be achieved
by preparing auxiliary entangled states, performing local
measurements and applying single-qubit operations. This approach
is at the heart of linear-optical quantum computing~\cite{Knill01}
and plays an important role in fault-tolerant quantum computation.
Basically, one prepares certain gates offline as the entanglement
resource for a teleportation protocol~\cite{NielsenChuang}. This
resource is then also easier to implement fault-tolerantly.
Experimentally, two-qubit gates have been teleported~\cite{Gao10}.

More generally, such gate teleportation strategies lie at the
heart of cluster state quantum computing~\cite{Raussendorf01} and
other schemes for measurement-based quantum
computing~\cite{MBQC2}. Here a number of nodes (qubits or
qumodes~\cite{WeeRMP}) are prepared in a multi-partite entangled
state (e.g., a cluster state) or another suitable tensor network
state. Then, a suitable measurement on one node teleports its
state onto another node with the concurrent application of a
desired quantum gate~\cite{Raussendorf01,Nielsen04}. Large cluster
states can be generated with CV
systems~\cite{Menicucci06,ZhangJ06}, with an experiment
implemented, involving more than $10^{4}$ optical
modes~\cite{Yokoyama13}.

\subsection{Port-Based Teleportation}

In port-based teleportation~\cite{Programmable,PortBased}
, Alice and Bob share $n$ 
Bell pairs, referred to as ports. Alice then performs a suitable
joint measurement on the system to be teleported and her $n$
qubits, communicating the outcome $k=1,\dots, n$ to Bob. Finally,
he discards all his systems except qubit $k$, which will be in the
unknown state of Alice's input. Compared to conventional
teleportation, port-based teleportation has the advantage that Bob
does not need to apply a correction at the end of the protocol.
The disadvantage is that the teleportation fidelity approaches
unity only in the limit of large $n$.

The importance of this scheme is less of a practical nature, due
to the large entanglement resources needed (even though this can
be improved by entanglement recycling~\cite{Recycling}), but
rather in conceptual studies in quantum information theory. It is
an important primitive for devising programmable quantum
processors~\cite{Programmable}, tailored to store a unitary
transformation and then applying it to an arbitrary state. It has
also been used in ideas of instantaneous non-local quantum
computation~\cite{PositionBased}, in attacking schemes of
position-based cryptography~\cite{PositionBased}, and in
assessments of communication complexity tasks. In fact, advantages
in quantum communication complexity have generally been linked to
violations of Bell inequalities, with port-based teleportation
being the main proof tool \cite{Buhrman}.

\section{Experimental Status and Challenges}

We now discuss the main experimental achievements and challenges.
Ideally, an experiment of quantum teleportation is successful when
the following basic requirements are met:

\begin{itemize}
\item[(1)] The input state is arbitrary (within a suitable alphabet).

\item[(2)] A third party, say Victor, supplies the input state to Alice and
independently verifies Bob's output state (e.g., via quantum state
tomography or fidelity estimation).

\item[(3)] Alice performs a complete Bell detection, being able to
distinguish an entire basis of entangled states.

\item[(4)] Teleportation fidelity exceeds the appropriate
threshold, achievable by classical measure-prepare strategies.
\end{itemize}

In the laboratory though, some of these conditions may not be met.
In particular, the failure of condition~(3) has led to
post-selected (or probabilistic) teleportation, where only a
subset of the Bell states are accessible. In this case, the
protocol has an associated Bell-efficiency, which provides an
upper bound to its overall success probability. By contrast, if
condition~(3) can be met, then one can in principle perform
unconditional (or deterministic) teleportation.

The most complete experiments of unconditional teleportation are
those with `real-time' (or `active') feed-forward, meaning that
the outcomes of the Bell measurement are communicated in real-time
to Bob, and the conditional unitaries are actively performed on
the output state, before being verified by Victor. In other
experiments the feed-forward is instead simulated in the
post-processing, i.e., the output state is directly measured by
Victor and the unitaries are simulated only after the tomographic
reconstruction of the state.

\begin{figure*}[t!]
\begin{center}
\includegraphics[width=1\textwidth] {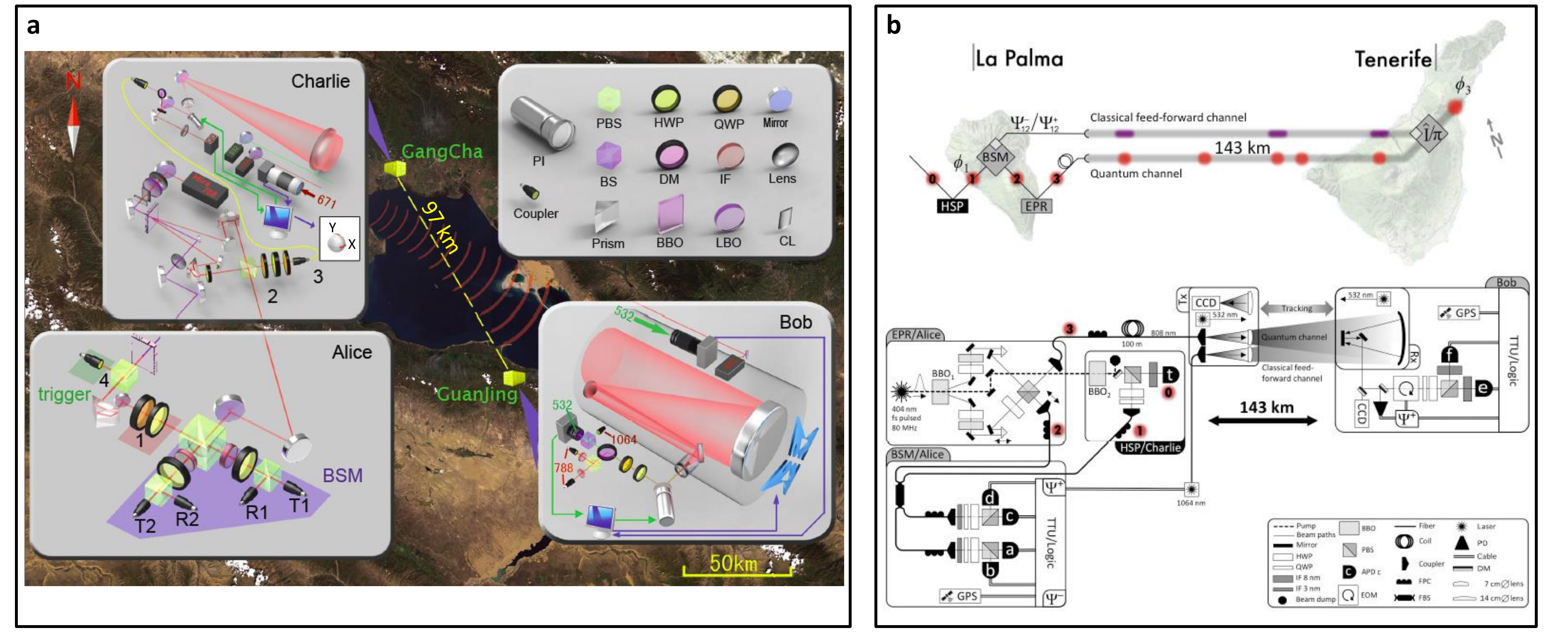}
\end{center}
\par
\vspace{-0.4cm} \caption{Long-distance quantum teleportation with
polarisation qubits. The experiment~\cite{Yin12} in
panel~\textbf{a} considered a free-space link of $\simeq 100~$km
at ground level across the Qinghai lake in China. Teleportation
was implemented with $50\%$ Bell-efficiency and a fidelity
$\gtrsim80\%$. The channel attenuation between Alice and Bob was
variable between 35 and 53~dB. Besides atmospheric loss, a major
attenuation came from the beam spreading wider than the 40~cm
aperture of the receiver telescope. The experiment relied on a
point-and-track control system for the telescopes to compensate
atmospheric turbulence and ground settlement. Entanglement
distribution over two $\simeq50$~km links (for a total of
$\simeq$80~dB loss) was also implemented. The
experiment~\cite{Ma12} in panel~\textbf{b} considered a free-space
link of 143~km at 2,400~m altitude between the Canary islands of
La Palma and Tenerife. Teleportation was implemented with $50\%$
Bell-efficiency (with active feed-forward) and a fidelity
$\gtrsim83\%$. A telescope with 1~m aperture was used at Bob's
site, and the total channel attenuation varied between 28 and
39~dB. Besides a tracking system, the experiment involved
techniques to enhance the signal-to-noise ratio, including
low-noise single-photon detectors with large active areas and
entanglement-assisted clock synchronisation. Figures adapted with
permission from: \textbf{a}, ref.~\protect\onlinecite{Yin12},
\copyright ~2012 NPG; \textbf{b}, ref.~\protect\onlinecite{Ma12},
\copyright ~2012 NPG.}\label{picLONG}
\end{figure*}

\subsection{Photonic Qubits}

The practical realisation of a complete Bell detection is still a
major issue for photonic qubits, since the simplest use of linear
optics and photodetection allows one to distinguish at most two of
the four Bell states~\cite{Weinfurter94,
Braunstein95,Calsamiglia01}, thus limiting the Bell-efficiency to
$50\%$. In principle, by using linear optics and $n$ ancillary
qubits, one can teleport one qubit with better
Bell-efficiency~\cite{Knill01}, approaching $100\%$ for infinite
$n$. This theoretical possibility clearly implies an overhead of
quantum resources which poses non-trivial experimental challenges.
A similar improvement can be achieved via quantum error
correction. One can build an $n$-qubit code correcting $2n-1$
localised erasures. Encoding the input qubit in such a code would
then protect against the inconclusive Bell outcomes, tolerating a
probability of erasure approaching $50\%$ for large $n$. Improving
the practical efficiency of the Bell detection with photonic
qubits is today an active area of
research~\cite{Grice11,Zaidi13,Ewert14}.



The Innsbruck experiment~\cite{Bouwmeester97} realised
post-selected teleportation of a polarisation qubit with $25\%$
Bell-efficiency, a performance later improved to $50\%$ in a 600~m
fibre-optical implementation across the river
Danube~\cite{Ursin04}. A point raised about the Innsbruck
experiment was that, since its implementation required Bob's
teleported qubit to be detected, condition~(2) above was rendered
problematic. One might argue that for this experiment
teleportation was achieved as a post-diction and was therefore
termed ``a posteriori teleportation''~\cite{Braunstein98b}.

To achieve complete Bell detection, the Rome
experiment~\cite{Boschi98} entangled photons in both spatial and
polarisation degrees of freedom. The input state was prepared
within the setup over one of the entangled photons, therefore
losing the ability to teleport entangled or mixed states. This
implies that the input state cannot be independently supplied from
outside, thus violating condition~(2) above.
The scheme was later extended to free-space teleportation over
16~km with real-time feed-forward and achieved $89\%$
fidelity~\cite{Jin10}. Complete Bell detection has also been
attempted using nonlinear interactions. Unfortunately, this leads
to an extremely inefficient protocol. Using optical
upconversion~\cite{Kim01}, only one out of roughly $10^{10}$
polarisation qubits takes part in the non-linear process.

A photonic qubit has also be realised as a time-bin
qubit~\cite{Brendel99}, i.e., as a single photon populating one of
two temporal modes. Time-bin qubits have been teleported along a
$2$~km telecom fibre with $25\%$ Bell-efficiency and $\simeq81\%$
fidelity~\cite{Marcikic03}. The qubit was teleported between
photons of different wavelengths (from $1.3~\mu $m to
$1.55~\mu$m). This experiment was extended to a relay
configuration~\cite{deRiedmatten04}, where the Bell detection was
performed by a third party, connected with Alice's input state and
the entangled source by means of 2~km fibres. Such a setup has the
challenge to preserve the indistinguishability of the photons
involved in the Bell detection. See ref.~\onlinecite{Landry07} for
an implementation on a commercial telecom network.

Photonic qubits are an excellent substrate for achieving long
distances. This is particularly true in free space where
birefringence is weak and photon absorption is small at optical
frequencies (e.g., compared to fibres). Two recent experiments
with polarisation qubits in free space~\cite{Yin12,Ma12} achieved
very long distances, suggesting that the atmosphere can be
traversed, and ground-to-satellite implementations are within
reach of current technology. See Fig.~\ref{picLONG} for more
details.

A complementary challenge is the photonic miniaturisation of
teleportation for applications in quantum processors based on
linear optics and probabilistic gates~\cite{Knill01}.
Ref.~\onlinecite{Metcalf14} teleported a dual-rail qubit (a single
photon encoded in one of two spatially-separated modes) within a
configurable photonic chip. Bell detection was performed with an
efficiency of $1/27$, and a fidelity of $89\%$ was extrapolated by
simulating the feed-forward operations in the post-processing,
after the tomographic reconstruction of Bob's output. Active
feed-forward remains a major challenge on photonic chips due to
the need for ultrafast photon detection and integrated electronics
operating with terahertz bandwidths.

Finally, we have witnessed the teleportation of multiple degrees
of freedom of a composite quantum system. Ref.~\onlinecite{Wang15}
simultaneously teleported the spin (polarisation) and orbital
angular momentum of a single photon, with a fidelity ranging from
$57\%$ to $68\%$. This is above the classical threshold of $40\%$
given by the optimal state estimation on a single copy of a
two-qubit system. Besides a hyper-entangled source, the experiment
exploited a hyper-entangled Bell detection with efficiency of
$1/32$. To measure the two degrees of freedom, the Bell detection
was designed as a quantum non-demolition measurement involving an
additional teleportation setup. 

\subsection{Nuclear Magnetic Resonance}

Complete Bell detection is achievable in NMR teleportation, where
the qubits are nuclear spins. Ref.~\onlinecite{Nielsen98}
considered the spins of two Carbon nuclei (C$_{1}$ and C$_{2}$)
and that of a Hydrogen nucleus (H) in a molecule of labelled
trichloroethylene (more precisely, an ensemble of such molecules,
so that the results are intended to be averaged). After
entanglement between C$_{1}$ and H was created by spin-spin
interactions, teleportation was realised from C$_{2}$ to H, by
making a complete Bell detection on the two Carbon
nuclei~\cite{Nielsen98}. This detection was performed in two
steps~\cite{Brassard98}: The Bell basis was first rotated to the
computational basis by radio-frequency (RF) pulses and a spin-spin
coupling; then, a
projection 
on this basis was done by exploiting the natural phase decoherence
of the Carbon nuclei (the fastest in the molecule, with times
$\leq0.4$~s). Since NMR teleportation is limited to inter-atomic
distances, it is not suitable for quantum communications, but
could still be used as a
subroutine for NMR quantum computing. 

\begin{figure*}[hptb]
\begin{center}
\includegraphics[width=1\textwidth] {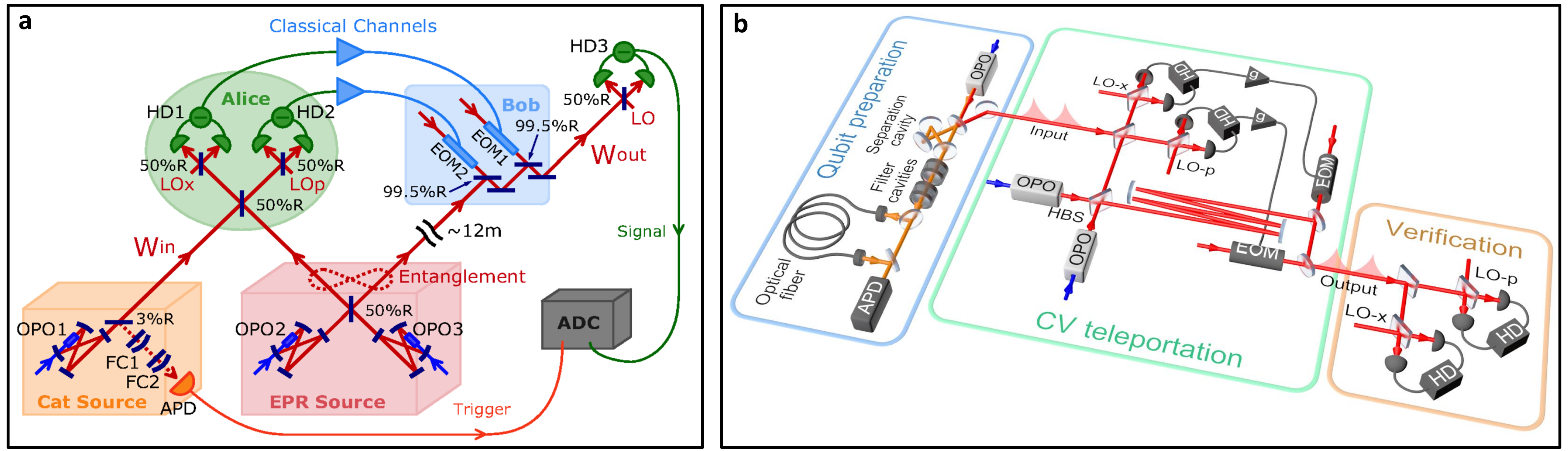}
\end{center}
\par
\vspace{-0.5cm} \caption{Quantum teleportation with optical modes.
Panel~\textbf{a} is the setup of a full CV teleportation
experiment with optical modes~\protect\cite{Lee11}. Here an EPR
state is generated by using two optical parametric oscillators
(OPOs) and a balanced beamsplitter. At Alice's station, part of
the EPR state is combined with the input state (here a cat state)
in a CV Bell detection, i.e., another balanced beamsplitter with
two homodyne detectors at the output ports. The classical outcomes
are transmitted to Bob who displaces his part of the EPR state by
using two EOMs. Note that both the preparation of the input state
and the verification of the output state are independently done by
Victor. Panel~\textbf{b} shows a hybrid teleportation experiment
where a time-bin qubit (prepared on the left) is teleported via CV
teleportation (the middle setup), involving an EPR state, a
complete CV Bell detection and EOMs~\protect\cite{Takeda13}.
Figures adapted with permission from: \textbf{a},
ref.~\protect\onlinecite{Lee11}, \copyright ~2011 AAAS;
\textbf{b}, ref.~\protect\onlinecite{Takeda13}, \copyright ~2013
NPG.}\label{picCV} \vskip -0.1truein
\end{figure*}

\subsection{Optical Modes}

The first and simplest resolution of the Bell detection problem
came from the use of CV systems, in particular, optical
modes~\cite{Furusawa98}. As previously mentioned, the CV version
of this measurement can easily be implemented with linear optics,
using a balanced beamsplitter followed by two conjugate homodyne
detections at the output ports. Both the beamsplitter visibility
and the homodyne quantum efficiency can be extremely high, so that
the overall detection can approach $100\%$ efficiency. CV
teleportation satisfies all the conditions (1)-(4) stated above,
with the only restriction that the input states must belong to a
specific alphabet, as a consequence of the infinite-dimensionality
of the Hilbert space. The first experiment with optical
modes~\cite{Furusawa98} considered an alphabet of fixed-energy
phase-modulated coherent states, achieving a fidelity of
$\simeq58\%$, above the relevant classical threshold.

This experiment has later been improved to beat the no-cloning
bound, reaching a fidelity of $\simeq70\%$ in
ref.~\onlinecite{Takei05} and $\simeq76\%$ in
ref.~\onlinecite{Yonezawa07}. These high-fidelity CV teleporters
are able to transfer nonclassical features such as non-positive
Wigner functions. Other experiments with coherent states involved
phase scanning at different amplitudes (ref.~\onlinecite{Zhang03}
with fidelity $\simeq61\%$) and simultaneous amplitude and phase
modulation (ref.~\onlinecite{Bowen03} with fidelity $\simeq
64\%$). Besides coherent states, other classes of states have been
teleported, including squeezed states~\cite{Takei05b,Yonezawa07},
Schr\"{o}dinger cat states~\cite{Lee11} and entangled states (see
entanglement swapping~\cite{Takei05,Jia04,Takeda15}). For typical
setups see Fig.~\ref{picCV}.

Unfortunately, CV teleportation fidelities cannot reach $100\%$,
since CV systems do not allow a maximally entangled state to be
prepared with finite resources. Realistic EPR states, with good
but finite squeezing, can be realised by producing two single-mode
squeezed states via parametric down-conversion (e.g.,
in optical parametric oscillators) and combining them 
on a balanced beamsplitter. Even with the high levels of squeezing
currently feasible ($\simeq$10dB), this resource also requires
good phase locking in a teleportation experiment. The current
record for CVs is a fidelity of $\simeq83\%$~\cite{Yukawa08}.

In order to overcome the limitations due to finite squeezing, CV
teleportation has been `integrated' with DVs into a hybrid
formulation~\cite{Furusawa11, Andersen13}.
Ref.~\onlinecite{Takeda13} realised the unconditional CV quantum
teleportation of a photonic qubit, by applying a broadband CV
teleporter to the temporal modes of a
time-bin qubit. 
Such an approach aims to simultaneously solve the problem of a
complete Bell detection (thanks to CVs) as well as achieving a
high teleportation fidelity thanks to DVs. Using just moderate
levels of squeezing, one can achieve fidelities $\gtrsim80\%$. 
 See also Fig.~\ref{picCV}(b).

Optical modes are well suited for integration into communication
technologies, thanks to the use of high-performance homodyne
detectors and off-the-shelf parts, like the electro-optical
modulators (EOMs) used for the input preparation and the output
displacements. While it is unclear whether the CV teleportation
distance is limited only to middle-range distances due to an
increased fragility with respect to loss (all previous experiments
have been table-top), the possibility to go broadband and achieve
high-rates is clearly appealing.

\subsection{Atomic Ensembles}

Quantum teleportation may be performed between quantum systems of
a different nature, for example, between light and matter.
Ref.~\onlinecite{Sherson06} was the first experiment to show this
feature by teleporting coherent states of an optical mode onto the
collective spin of an atomic ensemble, composed of around
10$^{12}$ room temperature Caesium atoms. Since the collective
transverse atomic spin components can be described by quadrature
operators
, this scheme implements a CV teleportation scheme.

\begin{figure*}[hptb]
\begin{center}
\includegraphics[width=0.96\textwidth] {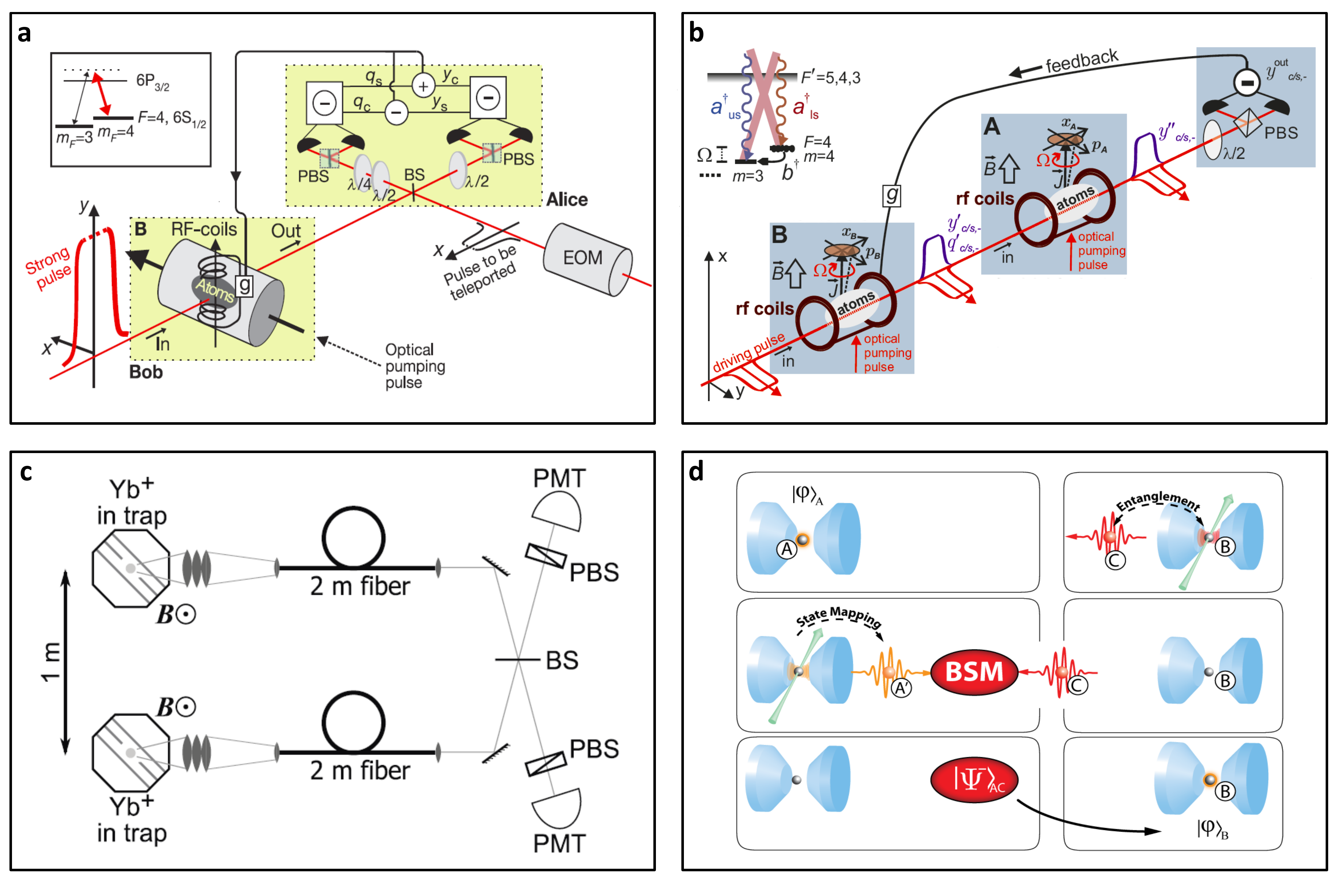}
\end{center}
\par
\vspace{-0.5cm} \caption{Quantum teleportation with matter. Upper
panels~(\textbf{a-b}) show teleportation experiments with atomic
ensembles of Caesium atoms at room temperature. Panel~\textbf{a}
shows the light-to-matter teleportation of a coherent state of an
optical mode into a collective atomic
spin~\protect\cite{Sherson06}, while panel~\textbf{b} shows the
matter-to-matter teleportation between the collective spins of two
atomic ensembles~\protect\cite{Krauter13} (see text for more
details). Lower panels~(\textbf{c-d}) show setups with single
trapped atoms, where teleportation is achieved by an incomplete
Bell detection on emitted photonic qubits. Panel~\textbf{c} shows
the teleportation between trapped Ytterbium ions, confined in
independent Paul traps~\protect\cite{Olmschenk09}, separated by
$1~$m. Each atomic qubit was encoded in the hyperfine states of
the $^{2}S_{1/2}$ level having coherence times of $\simeq2.5$~s,
and were manipulated by resonant microwave pulses at
$\simeq12.6~$GHz. Spontaneously emitted photons were coupled into
fibres, and directed to a beamsplitter, where the output ports
were filtered in polarisation and detected by single-photon PMTs
(implementing a Bell detection with $25\%$ efficiency).
Panel~\textbf{d} shows the steps of teleportation between two
neutral Rubidium atoms trapped in optical cavities separated by a
$21$~m optical fibre~\protect\cite{Nolleke13}. Qubits were encoded
in two Zeeman states of the atomic ground-state manifold.
Atom-photon entanglement and state mapping were efficiently
achieved by vacuum-stimulated Raman adiabatic passage. The
generated photonic qubits were then subject to an incomplete Bell
detection ($25\%$ efficiency). Figures adapted with permission
from: \textbf{a}, ref.~\protect\onlinecite{Sherson06}, \copyright
~2006 NPG; \textbf{b}, ref.~\protect\onlinecite{Krauter13},
\copyright~2013 NPG; \textbf{c},
ref.~\protect\onlinecite{Olmschenk09}, \copyright ~2009 AAAS;
\textbf{d}, ref.~\protect\onlinecite{Nolleke13}, \copyright ~2013
APS.}\label{picENS} \vskip -0.1truein
\end{figure*}

Light-matter entanglement was generated by sending a strong light
pulse through Bob's atomic sample\cite{Sherson06}, yielding
coherent scattering with the light being subject to the Faraday
effect and the atoms evolving by the dynamic Stark effect. The
optical output was combined with another optical pulse (modulated
by an EOM) at Alice's station, where a complete Bell measurement
was performed using a balanced beamsplitter followed by two sets
of polarisation homodyne detectors. Conditioned on the outcome,
Bob performed spin rotations on the atoms applying RF magnetic
field pulses. See Fig.~\ref{picENS}(a).

In this way, unconditional light-to-matter teleportation was
achieved~\cite{Sherson06} with a fidelity $\gtrsim58\%$. All the
conditions (1)-(4) stated above were achieved, for an input
alphabet of coherent states. We note that atomic ensembles have
the potential for good quantum memories~\cite{Simon10} with
sub-second coherence times and millisecond storage
times~\cite{Julsgaard04}. This is an important feature for the
construction of a scalable quantum network where teleportation is
used to store flying data (optical modes) into stationary media
(atomic ensembles).

This experiment was extended to unconditional matter-to-matter CV
teleportation~\cite{Krauter13}, between two ensembles of around
10$^{12}$ room temperature Caesium atoms, by using a four-wave
mixing interaction. As before, Bob's ensemble (B) was subject to a
strong driving pulse whose scattering created a co-propagating
sideband field (C) entangled with its collective spin. This field
reached Alice's ensemble (A) prepared in a coherent state by RF
magnetic field pulses. The interaction with A led to a partial
mapping of its state onto C, which was then subject to
polarisation homodyning. Conditioned on the outcome, Bob rotated
the spin of B via RF magnetic field pulses completing the
teleportation from A to B with a fidelity $\gtrsim55\%$. This
deterministic protocol was also used to teleport time-evolving
spin states, thus realizing the first example of stroboscopic
teleportation. See Fig.~\ref{picENS}(b).


Light-to-matter teleportation has also been achieved with DVs by
using cold atomic ensembles\cite{Chen08}. This increased the
distance and fidelity but made the protocol probabilistic due to
the $50\%$ Bell efficiency. Ref.~\onlinecite{Chen08} teleported a
polarisation qubit over $7$~m in fibre onto a collective atomic
qubit, made from two cold ensembles of $\simeq10^{6}$ Rubidium
atoms. Teleportation was performed with $\simeq78\%$ fidelity, and
the output state was stored for $\simeq8~\mu$s. This protocol was
extended to matter-to-matter teleportation, between two cold
ensembles of $\simeq10^{8}$ Rubidium atoms (at $\simeq$100~$\mu$K)
connected by a 150~m optical fibre~\cite{Bao12}. A single
collective atomic excitation (spin wave) was teleported between
the ensembles with the aid of polarisation qubits. By Raman
scattering, Bob's spin wave qubit (B) was entangled with a
photonic qubit (B') travelling to Alice. Here, Alice's spin wave
qubit (A) was converted into another photonic qubit (A'). Finally,
the Bell detection of A' and B' teleported the state of A onto B.
A fidelity of $\simeq88\%$ with storage times of $\simeq129~\mu$s
was reported. Note that sub-second storage times are possible with
cold atomic ensembles~\cite{Radnaev10}.

\subsection{Trapped Atomic Qubits}

Unconditional qubit teleportation can be realised with trapped
ions~\cite{Barrett04,Riebe04,Riebe07}, with all conditions (1)-(4)
being satisfied. Even if the distances are very limited, due to
the short-range Coulomb interactions, the storage times in these
systems are very long. This makes them good quantum memories for
using teleportation as a subroutine in quantum computing.

Ref.~\onlinecite{Barrett04} considered three Beryllium ions
($^{9}$Be$^{+}$) confined in a segmented linear Paul trap, with an
inter-ion spacing of $3~\mu$m. Qubits were provided by
ground-state hyperfine levels, coupled via stimulated Raman
transitions. Resonance fluorescence was used for complete Bell
detection and, exploiting spin-echo pulses, a fidelity of
$\simeq78\%$ was achieved. Independently,
ref.~\onlinecite{Riebe04} considered three Calcium ions
($^{40}$Ca$^{+}$) in a linear Paul trap, with an inter-ion
distance of $5~\mu$m. An ion qubit was encoded in a superposition
of the $S_{1/2}$ ground state and the metastable $D_{5/2}$ state
(with lifetime $\simeq1.16$~s) and manipulated by laser pulses
tuned to the $729$~nm quadrupole transition. Complete Bell
detection was performed by observing resonance fluorescence on
photo-multiplier tubes (PMTs). Using re-phasing spin-echo pulses,
a fidelity of $\simeq75\%$ was reported with storage lifetimes of
$\simeq10$~ms. This second experiment was improved~\cite{Riebe07}
to achieve a fidelity of $\simeq83\%$.


To increase the distance beyond micrometers, one could use a
photonic-matter interface where photonic qubits are employed as
quantum carriers for state mapping and entanglement distribution,
though a practical linear-optics implementation would provide
incomplete Bell detection. This approach was followed by
ref.~\onlinecite{Olmschenk09}, which reported the probabilistic
teleportation between two Ytterbium ions ($^{171}$Yb$^{+}$) with
$\simeq90\%$ fidelity (see Fig.~\ref{picENS}(c) for other
details). This setup provides long coherence times but the
free-space interface suffers from inefficient photon collection.
An alternate solution may come from cavity quantum
electrodynamics, where collection efficiencies are enhanced by the
strong atom-cavity coupling, but at the expense of shorter coherence times. 
Following this idea, ref.~\onlinecite{Nolleke13} reported the
probabilistic teleportation between two neutral Rubidium atoms
($^{87}$Rb) trapped in two distant optical cavities, with a
fidelity $\simeq88\%$. See Fig.~\ref{picENS}(d) for details.

\begin{figure*}[t!]
\begin{center}
\includegraphics[width=0.99\textwidth] {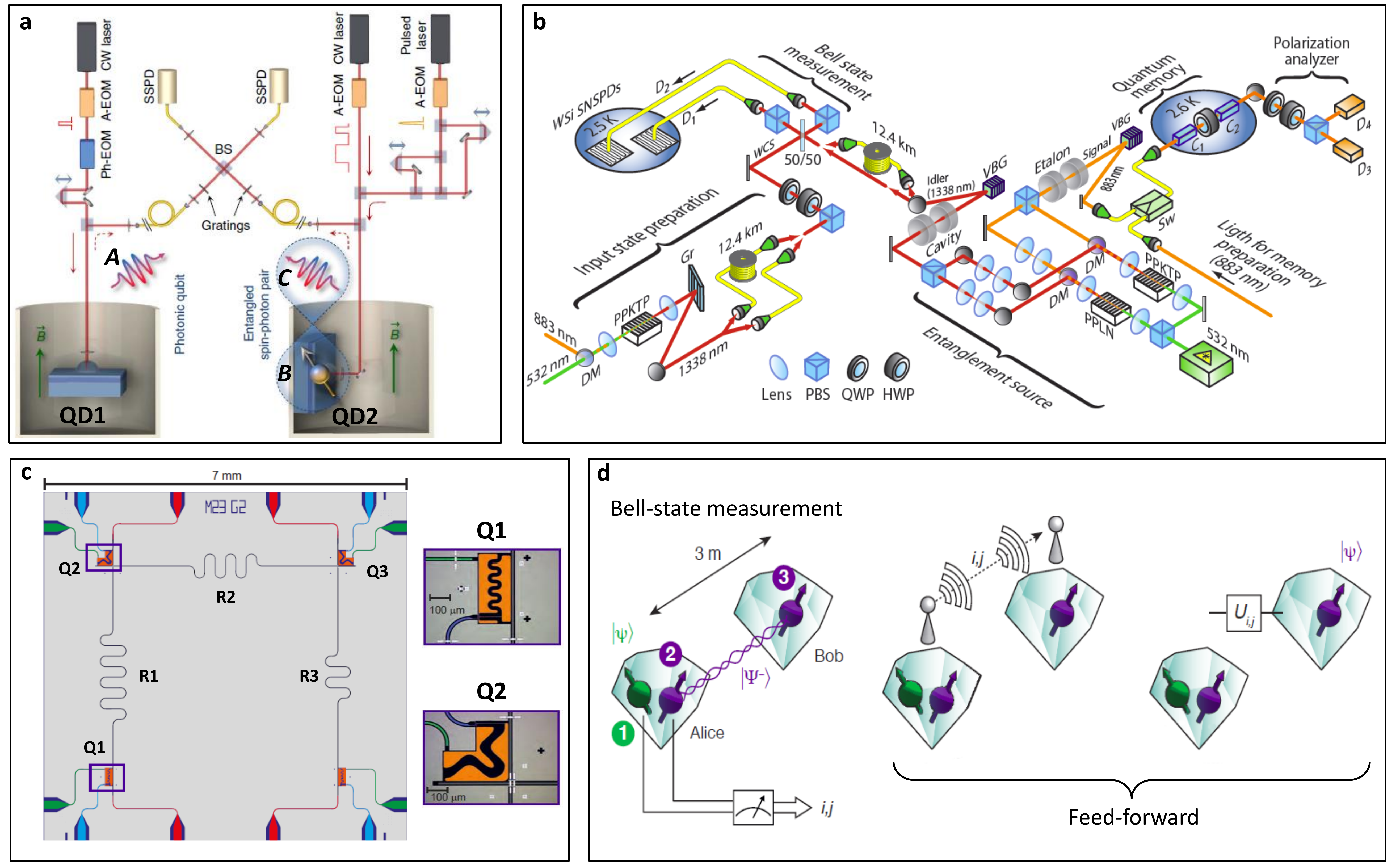}
\end{center}
\par
\vspace{-0.4cm}\caption{Quantum teleportation with solid state
systems. The setup in panel~\textbf{a} shows the teleportation of
a photonic frequency qubit $A$, generated by exciting a neutral
quantum dot (QD1 in the figure), onto the electron spin $B$ of a
charged quantum dot (QD2)~\protect\cite{Gao13}. By means of
resonant optical excitation, the spin $B$ was entangled with a
second frequency qubit $C$, sent to interfere
with $A$ in a Hong-Ou-Mandel interferometer. 
A coincidence detection at the output of the interferometer
heralded a successful photonic-to-spin teleportation with $25\%$
Bell efficiency. Panel~\textbf{b} shows the setup for teleporting
a telecom polarisation qubit onto a rare-earth doped crystal
quantum memory~\protect\cite{Bussieres14}. Two
polarisation-entangled photons were generated via spontaneous
parametric downconversion in nonlinear waveguides, and one of the
photons (at $883~$nm)
was stored in the quantum memory $10$~m away. The other photon (at $1338$%
~nm) was sent to interfere with the input polarised photon at the
Bell analyzer ($25\%$ Bell efficiency), where they were measured
by superconducting-nanowire single-photon detectors operating at
2.5~K. Teleportation of the specific state $\left\vert
+\right\rangle $ was extended to a total of $\simeq24.8~$km in
fibre. Panel~\textbf{c} shows the chip design for teleportation
between two superconducting transmon qubits, from Q1 to Q3, using
the middle qubit Q2 and the two resonators R1 and R2, which act as
quantum buses for entanglement generation and complete Bell
detection~\protect\cite{Steffen13}. The qubits were manipulated by
applying microwave pulses through individual charge gate lines,
and Bell state detection was performed with Josephson parametric
amplifiers. Panel~\textbf{d} shows the steps of teleportation
between two nitrogen-vacancy centres in
diamonds~\protect\cite{Pfaff14}. Once the electronic spins (2 and
3, in the figure) are entangled, the nitrogen nuclear spin (1) is
Bell-detected together with the electronic spin (2) of the same
centre. Finally, real-time feed-forward yields teleportation onto
the other electron spin (3). Figures adapted with permission from:
\textbf{a}, ref.~\protect\onlinecite{Gao13}, \copyright ~2013 NPG;
\textbf{b}, ref.~\protect\onlinecite{Bussieres14}, \copyright
~2014 NPG; \textbf{c}, ref.~\protect\onlinecite{Steffen13},
\copyright~2013 NPG; \textbf{d},
ref.~\protect\onlinecite{Pfaff14}, \copyright~2014
AAAS.}\label{picSOLID} \vskip -0.1truein
\end{figure*}

\subsection{Solid State Systems}

Probabilistic light-to-matter teleportation can be performed
between photonic qubits and solid-state quantum memories, composed
of semiconductor quantum dots (QDs)~\cite{Gao13} or rare-earth
doped crystals~\cite{Bussieres14}. The first
experiment~\cite{Gao13} probabilistically teleported a photonic
frequency qubit generated by a neutral QD (i.e., a single-photon
pulse in a superposition of two frequencies) onto the electron
spin of a charged QD separated by 5~m in a different cryostat. A
fidelity of $\simeq78\%$ was reported for the four equatorial
poles of the Bloch sphere (for which $F_{\text{class}}=3/4$) with
a spin coherence time extended to $\simeq13$~ns using spin-echo
pulses (see Fig.~\ref{picSOLID}(a) for details).

The second experiment~\cite{Bussieres14} probabilistically
teleported a telecom-wavelength polarisation qubit onto a single
collective excitation of a rare-earth doped crystal (a
neodymium-based quantum memory that stored photons for $50$~ns)
achieving a fidelity of $\simeq89\%$. One state was teleported
over long distance with slightly reduced fidelity, considering a
relay configuration where the Bell detection was performed at the
output of two 12.4~km long fibres (see Fig.~\ref{picSOLID}(b) for
details). This approach is promising for the availability of
quantum memories with long coherence times~\cite{Zhong15}.

Unconditional matter-to-matter teleportation has also been
achieved on solid state devices. 
Ref.~\onlinecite{Steffen13} considered superconducting transmon
qubits separated by $6$~mm and coupled to waveguide resonators in
a circuit quantum electrodynamics setup, as shown in
Fig.~\ref{picSOLID}(c). The experiment realised post-selected
teleportation (fidelity $\simeq81\%$), unconditional teleportation
(fidelity $\simeq77\%$), also including real-time feed-forward
(fidelity $\simeq69\%$) at a rate of 10$^{4}$Hz with microsecond
coherence times. The setup employed a quantum bus technology
scalable to more complex planar architectures, with non-trivial
implications for solid-state quantum computing.

Finally, ref.~\onlinecite{Pfaff14} teleported between two
nitrogen-vacancy centres in diamonds separated by 3~m, with
complete Bell detection and real-time feed-forward. These centres
were addressed by cryogenic confocal microscopes, and their spins
manipulated by spin-resolved optical excitations and microwave
pulses. The electronic spins of the centres were entangled via
entanglement swapping: coupling them with optical fields which
were subsequently overlapped at a beamsplitter and finally
detected. Upon success of this procedure, the nitrogen nuclear
spin of the first centre was prepared and Bell-detected together
with the electronic spin of the same centre. Teleportation onto
the electronic spin of the second centre was achieved with an
estimated fidelity of $\simeq86\%$. See Fig.~\ref{picSOLID}(d).
Nitrogen-vacancy centres are particularly attractive for combining
an optical interface (via the electron spin) with a long-lived
nuclear spin memory, having coherence times exceeding the second
at room temperature~\cite{Maurer12}.

\begin{table*}[t!]
\begin{center}
\begin{equation*}
\begin{tabular}{c|cccc}
\textbf{Quantum Technology} & \textbf{Efficiency} &
\textbf{Fidelity} & \textbf{Distance} & \textbf{Memory} \\ \hline
&  &  &  &  \\
\multicolumn{1}{r|}{Photonic qubits$~\left\{
\begin{array}{r}
\text{Polarisation~\cite{Bouwmeester97,Ursin04,Yin12,Ma12}} \\
\text{Time-bins}~\text{\cite{Marcikic03,deRiedmatten04,Landry07}} \\
\text{Dual-rails on chip~\cite{Metcalf14}} \\
\text{Spin-orbital qubits~\cite{Wang15}}
\end{array}
\right. $} & $
\begin{array}{c}
\leq 50\%^{\dagger} \\
25\% \\
1/27 \\
1/32
\end{array}
$ & $
\begin{array}{c}
\gtrsim 83\%~\text{\cite{Ma12}} \\
81\%^{\blacktriangle}\text{~\cite{Marcikic03}} \\
89\%^{\lozenge} \\
\gtrsim57\%^{\vartriangle}
\end{array}
$ & $
\begin{array}{c}
\text{143~km~\cite{Ma12}} \\
\text{6 km fibre~\cite{deRiedmatten04}} \\
\text{On chip} \\
\text{Table-top}
\end{array}
$ & $\left.
\begin{array}{c}
~ \\
~ \\
~ \\
~
\end{array}
\right\} $N/A$^{\ddagger}$ \\
&  &  &  &  \\
\multicolumn{1}{r|}{NMR~\cite{Nielsen98}} & $100\%$ & $\simeq
90\%^{\blacksquare}$ & $
\simeq $1~\AA  & $\simeq 1~$s \\
&  &  &  &  \\
\multicolumn{1}{r|}{Optical modes$~\left\{
\begin{array}{r}
\text{CVs~\cite{Furusawa98,Bowen03,Zhang03,Takei05,Takei05b,Yonezawa07,Lee11,Yukawa08}} \\
\text{Hybrid~\cite{Takeda13}}
\end{array}
\right. $} & $
\begin{array}{c}
100\% \\
100\%
\end{array}
$ & $
\begin{array}{c}
83\%\text{~\cite{Yukawa08}} \\
\gtrsim 80\%
\end{array}
$ & $
\begin{array}{c}
\text{12 m~\cite{Lee11}} \\
\text{Table-top}
\end{array}
$ & $\left.
\begin{array}{c}
~ \\
~%
\end{array}
\right\} $N/A$^{\ddagger}$ \\
&  &  &  &  \\
\multicolumn{1}{r|}{Atomic ensembles$~\left\{
\begin{array}{r}
\text{(hot) CV light-to-matter~\cite{Sherson06}} \\
\text{(hot) CV matter-to-matter~\cite{Krauter13}} \\
\text{(cold) DV light-to-matter~\cite{Chen08}} \\
\text{(cold) DV matter-to-matter~\cite{Bao12}}
\end{array}
\right. $} & $
\begin{array}{c}
100\% \\
100\% \\
50\% \\
50\%
\end{array}
$ & $
\begin{array}{c}
58\% \\
\gtrsim 55\% \\
78\% \\
88\%
\end{array}
$ & $
\begin{array}{c}
\text{Table-top} \\
0.5\text{~m} \\
7\text{~m fibre} \\
150\text{~m fibre}
\end{array}
$ & $
\begin{array}{c}
\left.
\begin{array}{c}
~ \\
~%
\end{array}
\right\} 4\text{ ms~\cite{Julsgaard04}} \\
\left.
\begin{array}{c}
~ \\
~%
\end{array}
\right\} 100\text{ ms~\cite{Radnaev10}}
\end{array}
$ \\
&  &  &  &  \\
\multicolumn{1}{r|}{Trapped atoms$~\left\{
\begin{array}{r}
\text{Trapped ions~\cite{Barrett04,Riebe04,Riebe07}} \\
\text{Trapped ions \& photonic carriers~\cite{Olmschenk09}} \\
\text{Neutral atoms in an optical cavity~\cite{Nolleke13}}
\end{array}
\right. $} & $
\begin{array}{c}
100\% \\
25\% \\
25\%
\end{array}
$ & $
\begin{array}{c}
83\%\text{~\cite{Riebe07}} \\
90\% \\
88\%
\end{array}
$ & $
\begin{array}{c}
5~\mu \text{m~\cite{Riebe04}} \\
1~\text{m} \\
21~\text{m fibre}
\end{array}
$ & $
\begin{array}{c}
\left.
\begin{array}{c}
~ \\
~%
\end{array}
\right\} \simeq 50\text{~s}^{\bigstar}\text{~\cite{Harty14}} \\
\begin{array}{c}
184~\mu \text{s~\cite{Specht11}}
\end{array}
\end{array}
$ \\
&  &  &  &  \\
\multicolumn{1}{r|}{Solid state$~\left\{
\begin{array}{r}
\text{Frequency qubit to quantum dot~\cite{Gao13}} \\
\text{Polarisation qubit to rare-earth crystal~\cite{Bussieres14}} \\
\text{Superconducting qubits on chip~\cite{Steffen13}} \\
\text{Nitrogen-vacancy centres in diamonds~\cite{Pfaff14}}
\end{array}
\right. $} & $
\begin{array}{c}
25\% \\
25\% \\
100\% \\
100\%
\end{array}
$ & $
\begin{array}{c}
78\%^{\S} \\
89\% \\
77\%\text{, }69\%^{\blacklozenge} \\
86\%
\end{array}
$ & $
\begin{array}{c}
5~\text{m} \\
\text{10~m, 24.8~km fibre}^{\square} \\
\text{On chip}~(6~\text{mm}) \\
3~\text{m}
\end{array}
$ & $
\begin{array}{c}
\gtrsim 1~\mu \text{s}^{\bigstar}\text{~\cite{Press10}} \\
1~ \text{ms}\text{~\cite{Jobez15}}, \simeq 6~\text{hours}^{\bigstar}\text{~\cite{Zhong15}} \\
\lesssim 100~\mu \text{s}^{\bigstar}\text{~\cite{XiangRMP13,KurizkiPNAS15}}\\
\simeq 0.6~\text{s}^{\triangledown}\text{~\cite{BarGill13}},
\gtrsim 1~\text{s}^{\blacktriangledown}\text{\cite{Maurer12}}
\end{array}
$
\end{tabular}
\end{equation*}
\caption{Comparison between quantum teleportation technologies in
terms of various quality factors, including the practical
efficiency of the Bell state analyzer, the maximum teleportation
fidelity, the maximum distance of teleportation in free-space (or
fibre, where explicitly indicated), and the storage time of
associated quantum memories. The storage time can be either the
typical decay time of the stored qubit (estimated from the
retrieval efficiency of the memory) or the time over which the
write-read fidelity drops below the classical threshold value
(e.g., this is the definition used for CVs). Coherence time is
shown in some cases. Notes: $\dagger$ Excluding setups with
internal qubit preparation~\cite{Boschi98,Jin10} for which
Bell-efficiency is $100\%$; $\ddagger$ Full optical delays based
on fibres or cavities have very short storage times;
$\blacktriangle$ A fidelity of 93\% was achieved for the four
equatorial states~\cite{Landry07}; $\lozenge$ Extrapolated
fidelity with feed-forward simulated in the post-processing;
$\vartriangle$ Two-qubit teleportation, whose corresponding
classical fidelity is $40\%$; $\blacksquare$ Entanglement
fidelity~\cite{Nielsen98}; $\bigstar$ Coherence time for this type
of quantum memory; $\S$ Equatorial states teleported; $\square$
One state teleported over 24.8~km in fibre with a fidelity of
$81\%$; $\blacklozenge$ Including real-time feed-forward;
$\triangledown$ Coherence time for electronic spins at low
temperatures ($\simeq77$~K); $\blacktriangledown$ Coherence time
for nuclear spins at room temperature.}
\end{center}\vskip -0.1truein
\end{table*}

\section{Discussion and Outlook}

We summarise the optimal experimental performances achievable by
various substrates and techniques in the table below. It is clear
that there is not yet an ideal quantum system or technology which
is able to excel in all parameters. Still, we may try to identify
preferred candidates in relation to specific applications of
quantum teleportation. Suitable hybridization of these candidates
with compatible substrates and techniques may provide the most
promising future developments for quantum teleportation and its
applications.

Short-distance teleportation ($\lesssim1$~m) as a quantum
computing subroutine is promising on solid-state devices, with the
best approach being circuit QED\cite{Schoelkopf08}. In particular,
superconducting transmon qubits may guarantee both deterministic
and high-fidelity teleportation on chip~\cite{Steffen13}. They
also allow the implementation of real-time feed-forward, which
appears to be challenging with photonic chips~\cite{Metcalf14}.
Moreover, they provide a more scalable architecture and better
integration to existing technology compared to earlier approaches
such as trapped-ions~\cite{Barrett04,Riebe04,Riebe07}. Currently
the only downside with these systems appears to be their limited
coherence time ($\lesssim100$~$\mu$s). This problem can be
overcome by integrating circuit QED with solid-state spin-ensemble
quantum memories (nitrogen-vacancy centres or rare-earth doped
crystals) which can provide very long coherence times for quantum
data storage. This is currently one of the biggest efforts of the
community~\cite{XiangRMP13,KurizkiPNAS15}.

Teleportation-empowered quantum communication at the metropolitan
scale (a few kms), could be developed using optical
modes~\cite{Furusawa98,Yukawa08}. For sufficiently low loss, these
systems provide high rates and bandwidths. Experiments may be
extended from table-top to medium-range implementations in fibre
or free-space and the potential integration with ensemble-based
quantum memories. Longer distances but lower rates may be achieved
via a hybrid approach~\cite{Takeda13} or by developing good
quantum repeaters based on non-Gaussian
operations~\cite{Eisert04}.

Long-distance quantum teleportation (beyond $100~$km) is an active
area but still affected by an open problem. Polarisation qubits
are the best carriers for (low-rate) teleportation over long
fibres and free-space links, but currently make the protocol
probabilistic due to the incomplete Bell detection. While
probabilistic teleportation and entanglement swapping are
acceptable for tasks such as entanglement distillation and quantum
cryptography, the situation is clearly different for quantum
communication where the input quantum information must be fully
preserved.

Accepting this probabilistic nature, satellite-based
implementations are within the reach of current technology.
Besides the integration of tracking techniques, the main challenge
comes from the high loss induced by beam spreading. This may be
overcome in a configuration where entanglement is distributed from
satellite to large-aperture ground-station telescopes. Assuming a
20~cm aperture satellite at $\simeq$600~km altitude and 1~m
aperture telescopes on the ground, one can estimate about $75~$dB
loss for a two-downlink channel, which is less than the
$\simeq80~$dB loss achieved at ground level for entanglement
distribution~\cite{Yin12}. Ground-to-satellite or inter-satellite
implementations are more challenging.

The future use of quantum teleportation as a building block of a
scalable quantum network strictly depends on its integration with
quantum memories. Such memories must come with excellent
radiation-matter interfaces, in terms of conversion efficiency,
write-read fidelity, storage time and bandwidth (high rate and
storage
capacity)~\cite{Lvovsky09,Simon10,Sangouard11,Bussieres13}. In the
first place, this would allow us to use quantum repeaters for
extending quantum communication well beyond direct transmission by
way of quantum error correcting codes. The development of good
quantum memories would enable not only the distribution of
entanglement across a network and quantum communication via
teleportation, but also the ability to coherently process the
stored quantum information. This could ultimately transform the
network into a world-wide distributed quantum computer or backbone
for a future quantum internet~\cite{Kimble2008}.

From this point of view, atomic ensembles are traditionally
considered appealing for their efficient light-to-matter
conversion~\cite{Simon10,Sangouard11} and their millisecond
storage times, which can approach the $\simeq $100~ms required by
light for transmission on a global scale. However, more promising
developments are today coming from solid state systems, where
excellent spin-ensemble quantum memories offer direct integration
with the scalable architecture of circuit QED. These memories may
not only extend the coherence time of circuit QED but also provide
an optical-microwave interface for inter-converting propagating
optical/telecom photons with on chip microwave
photons~\cite{OBrien14,KurizkiPNAS15}. Thus, a revised hybrid
architecture for the future Quantum Internet will likely be based
on long-distance quantum optical communication suitably interfaced
with solid-state nodes for quantum information processing.




\section{Acknowledgements}
S. P. has been supported by the Leverhulme Trust (qBIO fellowship)
and the EPSRC, via qDATA (Grant No. EP/L011298/1) and the UK
Quantum Communications Hub (Grant No. EP/M013472/1). J. E. has
been supported by BMBF (Q.com), the EU (SIQS), and the ERC (TAQ).
The authors would like to acknowledge useful feedback from U. L.
Andersen, G. Chiribella, N. Gisin, A. Imamoglu, C.-Y. Lu, P. van
Loock, S. Mancini, C. Monroe, S. Olmschenk, J. W. Pan, W. Pfaff,
E. Polzik, S. Popescu, T. C. Ralph, V. Scarani, C. Simon, R. Thew,
W. Tittel, A. Wallraff, and D. J. Wineland.

\end{document}